\documentclass[conference]{IEEEtran}
\IEEEoverridecommandlockouts
\usepackage{cite}
\usepackage{amsmath,amssymb,amsfonts}
\usepackage{algorithmic}
\usepackage{graphicx}
\usepackage{textcomp}
\usepackage{xcolor}
\usepackage{soul}
\usepackage{url}
\usepackage{comment}
\usepackage{siunitx}
\usepackage[font=small]{caption}
\usepackage{subfiles}
\usepackage{subcaption}
\usepackage{fancyhdr}
\usepackage[absolute,overlay]{textpos}
\usepackage{tabularx}
    \newcolumntype{L}{>{\raggedright\arraybackslash}X}
\usepackage{tikz}
\usetikzlibrary{arrows.meta,positioning,fit,shapes.multipart}
\def\BibTeX{{\rm B\kern-.05em{\sc i\kern-.025em b}\kern-.08em
    T\kern-.1667em\lower.7ex\hbox{E}\kern-.125emX}}

    \newcommand{\erfc}{\text{erfc}}
\begin{document}
\begin{textblock}{20}(1,0.5)  
    \noindent J. Wu, P. Sen, A. Singh, and S. Nie, ``Finite-Aperture Near-Field Terahertz Diffraction Modeling and Parametric Blockage Localization,'' \textit{to appear in Proc. IEEE 27th Int. Workshop Signal Process. Adv. Wireless Commun. (SPAWC)}, Athens, Greece, September 2026.
\end{textblock}

\title{Finite-Aperture Near-Field Terahertz Diffraction Modeling and Parametric Blockage Localization}

\author{\IEEEauthorblockN{\textbf{Junjie Wu}\IEEEauthorrefmark{1},
\textbf{Priyangshu Sen}\IEEEauthorrefmark{2}, 
\textbf{Arjun Singh}\IEEEauthorrefmark{2}, and
\textbf{Shuai Nie}\IEEEauthorrefmark{1}\IEEEauthorrefmark{3}} 
\IEEEauthorblockA{\IEEEauthorrefmark{1}School of Computing,
University of Nebraska-Lincoln, Lincoln, NE 68588 USA}
\IEEEauthorblockA{\IEEEauthorrefmark{2}SUNY Polytechnic Institute, Utica, NY 13502 USA}
\IEEEauthorblockA{\IEEEauthorrefmark{3}School of ECE,
University of Georgia, Athens, GA 30602 USA}
\IEEEauthorblockA{Emails: \url{jwu67@huskers.unl.edu}, \url{shuainie@uga.edu}}}

\maketitle

\begin{abstract}
Diffraction in terahertz (THz)-band near-field propagation channels with large apertures is not well analyzed in classical point-source knife-edge models. In this work, an analytical finite-aperture near-field model is developed based on the Rayleigh-Sommerfeld integral method. The model provides a closed-form expression for the received field with low complexity in evaluation. Results show that the received power after knife-edge diffraction is more sensitive to the edge height than to the edge's axial distance. This novel finding leads to a rotation-based parametric sensing method to estimate the location of the knife-edge blockage. The diffraction model is validated using COMSOL full-wave simulation and a 300-GHz testbed, while the localization framework is evaluated using extensive simulation. The work shows that the aperture dimension is critical in characterizing THz near-field diffraction and the received field can assist in low-complexity blockage localization.  

\end{abstract}


\section{Introduction}
\label{sec:intro}
As future generations of wireless systems are eyeing higher throughput and lower latency, terahertz (THz) spectrum, ranging from 100 GHz to 10 THz, has become a promising candidate due to the abundant underexplored bandwidth ~\cite{jornet2024evolution}. With the help of large antenna arrays with more than 100 elements to compensate for path losses caused by atmospheric attenuation and molecular absorption, THz communication can provide hundreds of Gbps links that can extend to nearly three kilometers~\cite{sen2023multi, o2024long}. However, the use of large antenna arrays inevitably expands the radiative near-field region where the classical plane-wave approximation no longer holds. In the near field, propagation of spherical wavefronts in free space does not follow the Friis formula~\cite{singh2025novel}. Similarly, near-field diffraction requires special treatment, yet a closed-form analytical model for the THz band remains elusive. 

In the meantime, emerging work in wavefront engineering and curved beam designs has further advanced near-field solutions that provide robust links at THz frequencies ~\cite{singh2023wavefront, guerboukha2024curving}. However, the beam design still requires certain determination on prior knowledge of the blockage, which remains a missing step in existing curved beam design. In other words, the ability to estimate the position of the unknown blockage will further strengthen the design of curved beams.

In this work, we first develop an analytical model based on the Rayleigh-Sommerfeld integral method to characterize the finite-aperture near-field THz diffraction, which is validated using both full-wave simulation and experiments. Based on the diffraction model, we establish a parametric blockage-sensing framework to estimate the location of the unknown blockage. 


\textbf{Related Work:} Diffraction has been widely studied in wireless propagation channel characterization since the early stage of wireless communication and radar systems~\cite{rappaport2024wireless}.
Classical diffraction models, such as the uniform theory of diffraction (UTD) and the stochastic theory of edge diffraction in~\cite{franceschetti2008stochastic}, provide a solution for far-field diffraction. In the near field, theoretical models are established for wideband scattering~\cite{chen2019near,dogaru2021near}. In leveraging the diffraction mechanism for sensing purposes, researchers in~\cite{pallaprolu2025embracing} utilize edge diffraction at sub-6 GHz and geometrical theory of diffraction to infer environmental information, such as surface geometries and material properties, of objects under study. At millimeter-wave frequencies, studies in~\cite{yu2022long} and~\cite{dizche2022early} analyze diffracted signal's characteristics for blockage prediction. Given the distinct propagation characteristics at THz frequencies, the method may fall short in capturing the unique features in the THz near-field domain. 

Summarizing the existing work, to the best of the authors' knowledge, little effort has been dedicated towards a communication-centric analytical knife-edge diffraction model in a finite-aperture near-field THz channel and the feasibility of blockage localization using the parametric inverse diffraction model. Bridging this gap will provide an environmental basis for more efficient curved beam design.
\vspace{-1mm}
\section{Near-Field Diffraction Model}
\label{sec:forwardModel}

Classical knife-edge diffraction (KED) models are typically derived under the assumptions of point-source illumination with a uniform field distribution~\cite{rappaport2024wireless}. In contrast, large-aperture THz systems operate in the near field, where the incident field is non-uniform across the aperture and its magnitude and phase vary along propagation. As a result, the classical model is not sufficient to describe diffraction in this regime, requiring a finite-aperture near-field diffraction model.

\subsection{Gaussian Beam Illumination}
Finite-aperture antennas naturally exhibit amplitude tapering from the center to the edge, resulting in a non-uniform illumination across the aperture and a varying beam profile along propagation, which can be well modeled by a Gaussian beam. The Gaussian beam is analogous to beamforming in the near-field~\cite{edmundopticsGaussianBeam}. 
The electric field of a Gaussian beam with a waist of $2\omega_0$ propagating along the $x$-axis follows the expression ~\cite{alda2003laser} 
\begin{equation}
E(x,r) = E_{0}\frac{\omega_0}{\omega(x)}e^{\left(\frac{-r^2}{\omega(x)^2}\right)} e^{\left( -i(kx + k\frac{r^2}{2R(x)} + \phi(x))\right)},
\label{eq:electric_field}
\end{equation} where $E_0$ is the electric field at the origin, $r$ is the radial distance from the center of the beam, $k$ is the wave number, $\phi(x)$ is the Gouy phase, and $R(x)$ is the radius of the beam's wavefront curvature. $\omega(x)$ is the beam waist at propagation distance $x$ and can be expressed as 
\begin{equation}
    \omega(x) = \omega_{0}\sqrt {1 + \left(\frac{x}{x_{R}}\right)^{2}},
\end{equation} 
where $x_R=\frac{\pi\omega_0^2}{\lambda}$ is the Rayleigh distance.

\begin{figure}[t]
\centering
\includegraphics[width=0.25\textwidth, trim=4mm 5mm 4mm 4mm, clip]{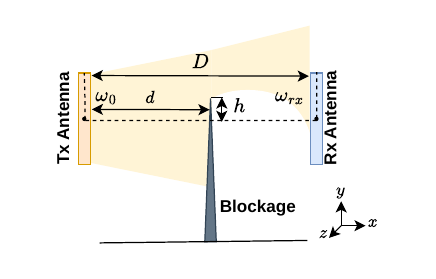}
\caption{System geometry used in the diffraction model. 
}
\vspace{-3mm}
\label{fig:geometry}
\end{figure}

As shown in Fig. \ref{fig:geometry}, the transmitter (Tx) is located at $x=0$ and the receiver (Rx) at $x=D$. The knife-edge blockage is assumed to be infinitely long along the $z$-axis and perpendicular to the $x$-axis, making the geometry invariant in the $z$ direction. 
The diffraction is therefore modeled in the $x$-$y$ plane, where the transmitted field distribution along the $y$-axis is expressed as: 
\begin{equation}
E(0,y)=E_0 e^ {\!-\frac{y^2}{\omega_{0}^2}}.
\label{eq:init_field}
\end{equation}

\vspace{-3mm}
\subsection{Rayleigh-Sommerfeld Integral Method} 
The 2D Rayleigh--Sommerfeld (RS) integral \cite{2D_RS}, which follows from Green's theorem and the Huygens principle, provides a rigorous expression for the diffracted field.
Each point on the source plane acts as a secondary source that radiates a cylindrical wave in the $x$-$y$ plane, and the total field at any observation point is obtained by superposing the contributions from all source points on the aperture.  

Let $P(x,y)$ denote an observation point and $Q(x',y')$ a source point on the Tx aperture plane. The distance between $Q$ and $P$ is
$R = \sqrt{(x-x')^2 + (y-y')^2}$. 
Therefore, the field at $P(x,y)$ is expressed as:
\begin{equation}
E(x,y) = \frac{1+i}{\sqrt{2\lambda}}
\int_{-\infty}^{\infty} E(x', y')\,
\frac{e^{-ikR}}{\sqrt{R}} \cos \theta \, dy'.
\label{eq:2D_RS}
\end{equation}

\subsection{Paraxial Approximation}
The Gaussian beam is a solution 
of the Helmholtz equation under the paraxial approximation\cite{alda2003laser}. Since the beam propagates with a small divergence angle around the propagation axis, the same approximation is applied to simplify \eqref{eq:2D_RS}. 
With $\Delta = x-x'$ denoting the axial distance, the obliquity factor satisfies $\cos \theta \approx 1$, and the propagation distance in the phase term can be approximated by:
\begin{equation}
    R = \sqrt{\Delta^2+(y-y')^2} \approx \Delta +\frac{(y-y')^2}{2\Delta}.
    \label{eq:approximation1}
\end{equation}
The amplitude term varies slowly along the $y$-axis, giving $\frac{1}{\sqrt{R}} \approx \frac{1}{\sqrt{\Delta}}$.
Applying these approximations to \eqref{eq:2D_RS} yields
\begin{equation}
E(x,y) = \frac{1}{\sqrt{-i\lambda}}
\int_{-\infty}^{\infty} E(x', y')\,
\frac{e^{-ik(\Delta+\frac{(y-y')^2}{2 \Delta })}}{\sqrt{\Delta}} \, dy'.
\label{eq:2D_RS_final}
\end{equation}

\subsection{Field at the Blockage Plane}
The blockage is located at $x=d$ and is assumed to be a perfect electric conductor. 
The vertical distance from the edge tip to the line-of-sight (LOS) path between the Tx and Rx centers is defined as the edge height $h$ along the $y$-axis.

When the Gaussian beam impinges on the knife edge, the field immediately before the edge at $x=d^-$ can be expressed by substituting the incident field in  \eqref{eq:init_field} into \eqref{eq:2D_RS_final} with $\Delta=d$:
\begin{equation}
    E( d^-,y)=E_0 e^{-ikd}  \frac{\omega_{0}}{\sqrt{Z_d}}   e^{\left( {-\frac{y^2}{Z_d}} \right)},
\label{eq:field_d-}
\end{equation}
where $Z_d=\omega_{0}^2-\frac{2id}{k}$ is the complex beam parameter at distance $d$. 
The edge blocks the field when $y<h$, which can be modeled using a gate function  
\begin{equation}
u(y-h)=
\begin{cases}
1,& y\ge h,\\
0,& y<h.
\end{cases}
\label{eq:mask}
\end{equation}
Therefore, the field immediately after the edge at $x=d^+$ is obtained by the element-wise product:
\begin{align}
    E(d^+,y)&=E(d^-,y)\times u(y-h)\\
    &= \begin{cases} 
E_0 e^{-ikd} \frac{\omega_{0}}{\sqrt{Z_d}} e^{\left( {-\frac{y^2}{Z_d}} \right)}, & y \ge h, \\
0, & y < h.
\end{cases}
\label{eq:E_d+}
\end{align}
\subsection{Field and Power at Receiver Plane}
After the edge, the remaining wavefront continues to propagate. Substituting \eqref{eq:E_d+} into \eqref{eq:2D_RS_final} with $\Delta=D-d$, the received field at $x=D$ becomes
\begin{equation}
    E(D,y) = \frac{E_0 \omega_{0} e^{-ikD}}{2\sqrt{Z_D}} e^ {\left( -\frac{y^2}{Z_D} \right)} \erfc \left( \Phi(y) \right),
    \label{eq:E(D,y)}
\end{equation}
where $Z_D = \omega_{0}^2 - \frac{2iD}{k}$ and $\Phi(y)=\frac{Z_D h - y Z_d}{\sqrt{Z_d Z_D (Z_D - Z_d)}}$. The total received power $P_r$ is obtained by integrating the field intensity $E(D,y)$ over the receiver aperture $[-\omega_{r}, \omega_{r}]$,
\begin{equation}
P_r = \frac{1}{2\eta} \int_{-\omega_{r}}^{\omega_{r}} |E(D,y)|^2 \, dy,
\label{eq:Pr_cal}
\end{equation}
where $\eta$ is the intrinsic impedance of free space.
Substituting \eqref{eq:E(D,y)} into \eqref{eq:Pr_cal} yields:
\begin{equation}
P_r =\frac{|E_0|^2 \omega_{0}^2}{8\eta |Z_D|}
 \int_{-\omega_{r}}^{\omega_{r}} e^ {\left( -\frac{2y^2 \omega_{0}^2}{|Z_D|^2} \right)} \left | \erfc \left( \Phi(y) \right) \right|^2 dy,
 \label{eq:Pr}
\end{equation}
which expresses the received power as a function of the blockage location. 

\begin{figure}[t]
    \centering
    \begin{subfigure}{0.48\linewidth}
        \centering
        \includegraphics[width=\linewidth, trim=0 0 0 0, clip]{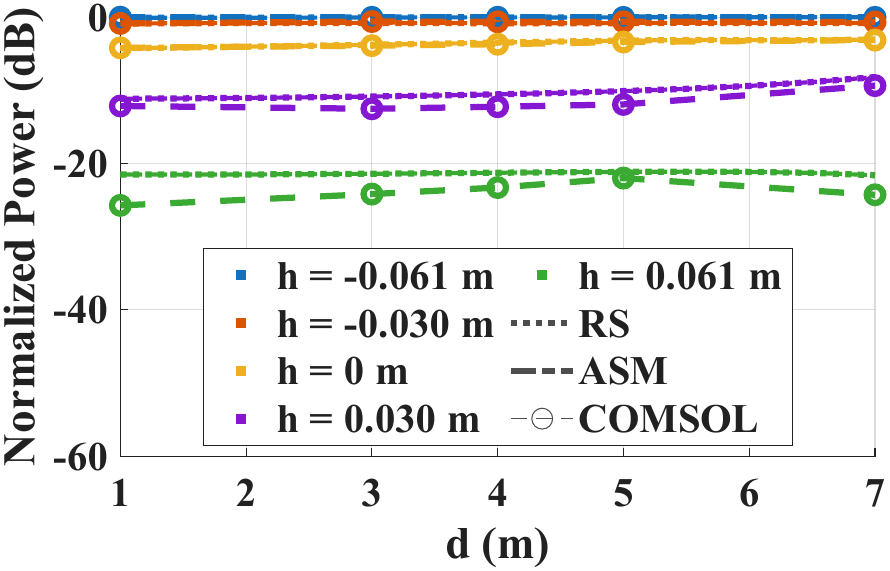}
        \caption{Normalized power vs. $d$.}
        \label{fig:Pr_d}
    \end{subfigure}
    \hfill
    \begin{subfigure}{0.48\linewidth}
        \centering
        \includegraphics[width=\linewidth, trim=0 0 0 0, clip]{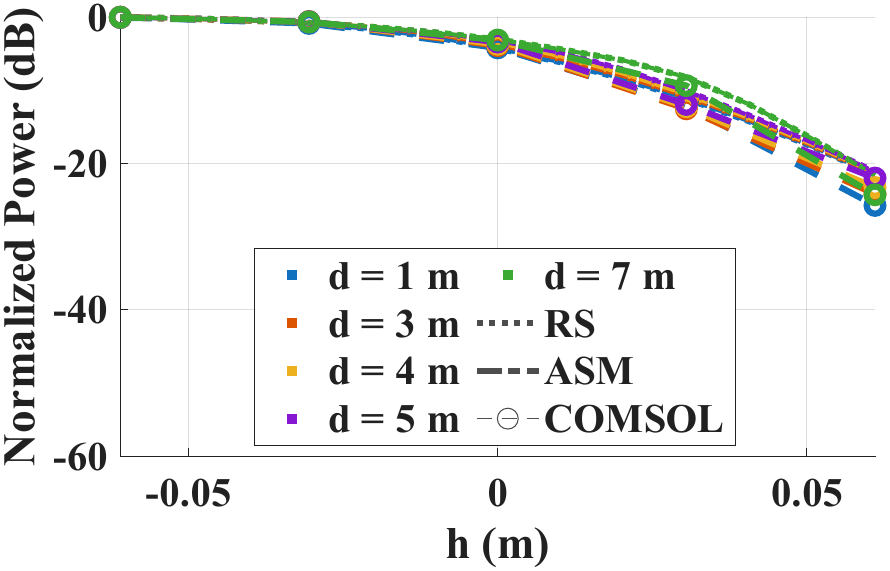}
        \caption{Normalized power vs. $h$.}
        \label{fig:Pr_h}
    \end{subfigure}
    \caption{Comparison of normalized received power from the RS integral, ASM, and COMSOL simulation versus (a) axial distance $d$ for different $h$ and (b) edge height $h$ for different $d$.}
    \vspace{-3mm}
    \label{fig:Pr_analysis_simulation}
\end{figure}

\subsection{Model Analysis}
\label{sec:forward_model_analysis}
In addition to the RS integral method, we also derive the model using the Angular Spectrum Method (ASM). The results from the RS and ASM formulations agree well with each other and with the COMSOL simulations, as shown in Fig.~\ref{fig:Pr_analysis_simulation}, thereby validating the proposed theoretical model for THz finite-aperture near-field diffraction. The COMSOL full-wave simulations were conducted using the Wave Optics Module, where the Tx and Rx were modeled as line boundaries.
Here, the received power is normalized with respect to the LOS received power for consistent comparison. A key observation is that the normalized power after diffraction exhibits a strongly asymmetric dependence on the edge location parameters $d$ and $h$. Specifically, it varies only weakly with $d$ but significantly with $h$, indicating that the received power primarily reflects the height of the edge. 

\begin{figure}[t]
    \centering
    \begin{subfigure}{0.5\linewidth}
        \centering
        \includegraphics[width=\linewidth, trim=0 0 0 0, clip]{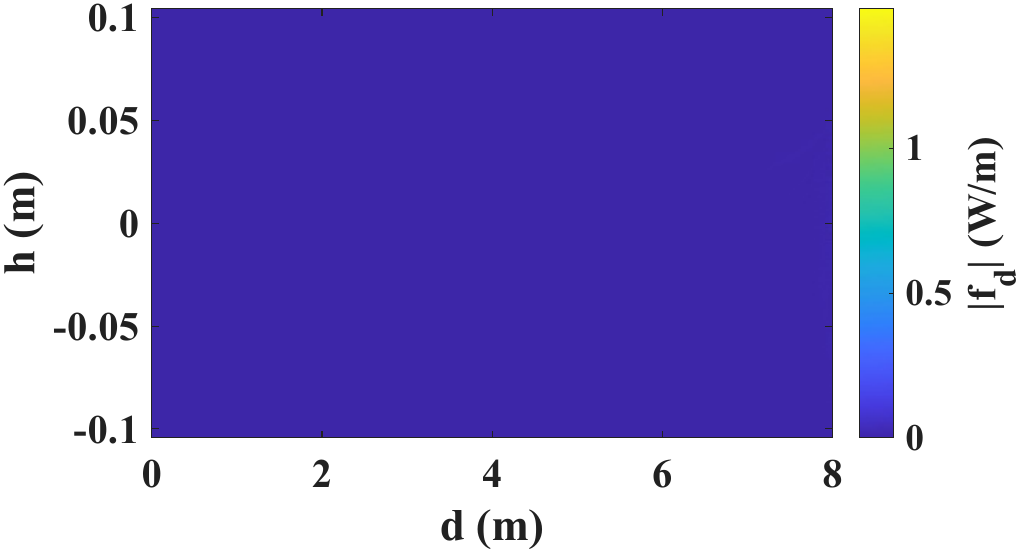}
        \caption{Sensitivity to $d$.}
        \label{fig:fd}
    \end{subfigure}\hfill
    \begin{subfigure}{0.5\linewidth}
        \centering
        \includegraphics[width=\linewidth, trim=0 0 0 0, clip]{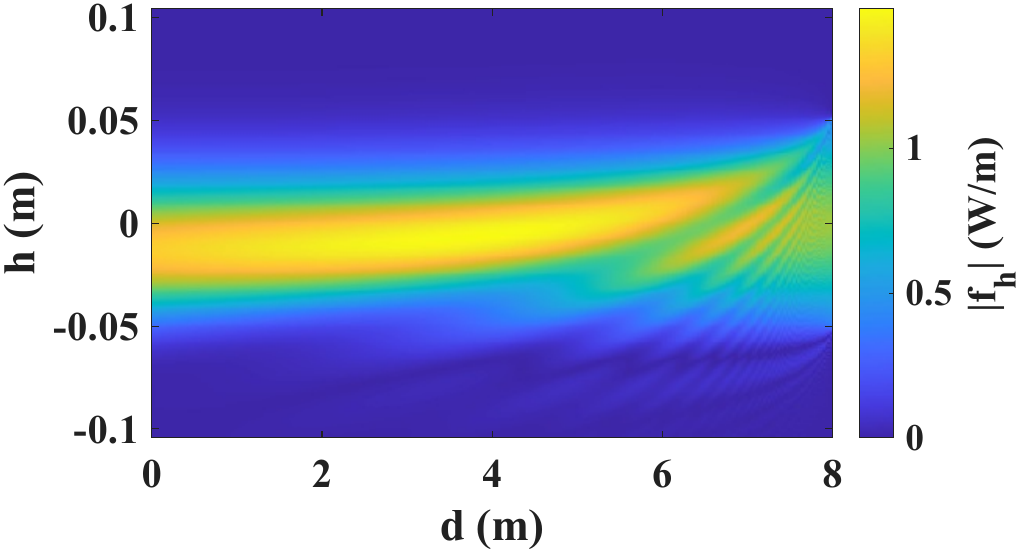}
        \caption{Sensitivity to $h$.}
        \label{fig:fh}
    \end{subfigure}
    \caption{Local sensitivity of the received power $P_r$ to (a) axial distance $d$ and (b) edge height $h$ at different blockage locations. 
    }
    \vspace{-3mm}
    \label{fig:sensitivity_analysis}
\end{figure}

To further quantify this behavior, we examine the local sensitivity of $P_r$ to $h$ and $d$, 
\begin{equation}
f_h = \left|\frac{\partial P_r}{\partial h}\right|, 
\qquad
f_d = \left|\frac{\partial P_r}{\partial d}\right|.
\end{equation}
The sensitivity maps shown in Fig. \ref{fig:sensitivity_analysis} 
confirm that $f_d\approx0$ and $f_h \gg f_d$ over the main beam region. Specifically, Fig.~\ref{fig:fd} shows that $f_d$ remains nearly zero over the considered range of $d$ and $h$, while Fig.~\ref{fig:fh} exhibits a high-sensitivity region of $f_h$ near the propagation axis.
The asymmetric sensitivity originates from the highly collimated beam generated by the large aperture in the THz near field. The beam spreads slowly along the propagation direction, so moving the edge along this direction makes minor changes in $P_r$. In contrast, varying $h$ directly changes the blocked portion of the beam, which leads to a much stronger effect. This motivates the design of a sensing framework to improve the observability of $d$ for blockage localization, as detailed in Sec. ~\ref{sec:sensing}.

\section{Experimental Validation} 
\label{sec:experiments}

We utilize the Advanced Communications Electronics and Sensing (ACES) Testbed~\cite{singh2025novel} established at SUNY Polytechnic Institute, as illustrated in Fig.~\ref{fig:ACES_setup}, for experimental validation of the diffraction model. At the transmitter side, an up-converter with a $\times 12$ multiplier and mixer chain from Virginia Diodes Inc. (VDI) operates within the J-Band (220--330 GHz). To generate the target RF signal at 300 GHz, the up-converter mixes inputs from an intermediate frequency (IF) at 2 GHz and a local oscillator (LO) signal at 25 GHz. The power amplifier offers 26 dB of gain, allowing a maximum output of 8.75 dBm before transmission from the antenna.

The receiving end has a similar architecture to the transmitter, featuring a VDI front end that down-converts the RF signal to 2 GHz along with a high-gain, low-noise amplifier (LNA) that provides 12 dB of gain at the IF. A digital version of the sampled down-converted signal is captured with a sampling rate of 50 GSa/s. This data is then processed by a signal processing backend using MATLAB to analyze the received power. We specifically use a 46-dBi lens antenna at both the transmitter and receiver, with an aperture size of $12.2$ cm, and a resultant near-field of 25~m at 300~GHz to closely align with the simulation model.

\subsection{Experimental Methodology}
\begin{figure}[t]
    \centering
    \includegraphics[width=1\linewidth]{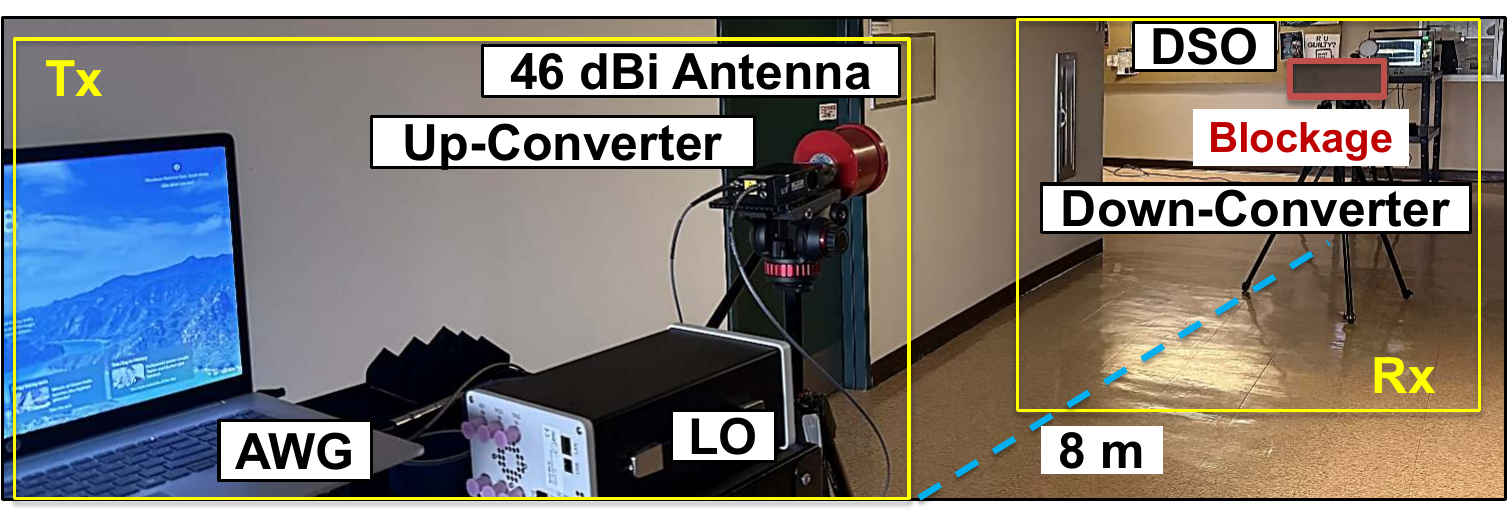}
    \caption{ACES testbed setup.}
    \vspace{-4mm}
    \label{fig:ACES_setup}
\end{figure}

The experimental measurements were conducted in a hallway where the walls are primarily made of drywall, as shown in Fig.~\ref{fig:ACES_setup}, while the antennas are aligned in the broadside orientation. We established a LOS link with the transmitter and receiver located $8$ meters apart and introduced a knife-edge blockage using a $16\times20$~cm rectangular metal plate at various distances from $1$ to $8$ meters, in 1-meter increments. To assess the extent of the blockage, we gradually adjusted the vertical position of the metal plate in relation to the fixed antenna aperture using the following measurement mechanism: i) at $-6.1$ cm, the blockage is at the bottom edge of the antenna aperture, causing no obstruction to the aperture, ii) at $-3.05$ cm, there is a quarter blockage, iii) at $0$ cm, there is a half blockage, iv) at $3.05$ cm, there is a three-quarters blockage and v) at $6.1$ cm, there is full blockage of the antenna aperture.

\subsection{Experimental Results}
\begin{figure}[t]
    \centering
    \begin{subfigure}{0.48\linewidth}
        \centering
        \includegraphics[width=\linewidth, trim=0 0 0 0, clip]{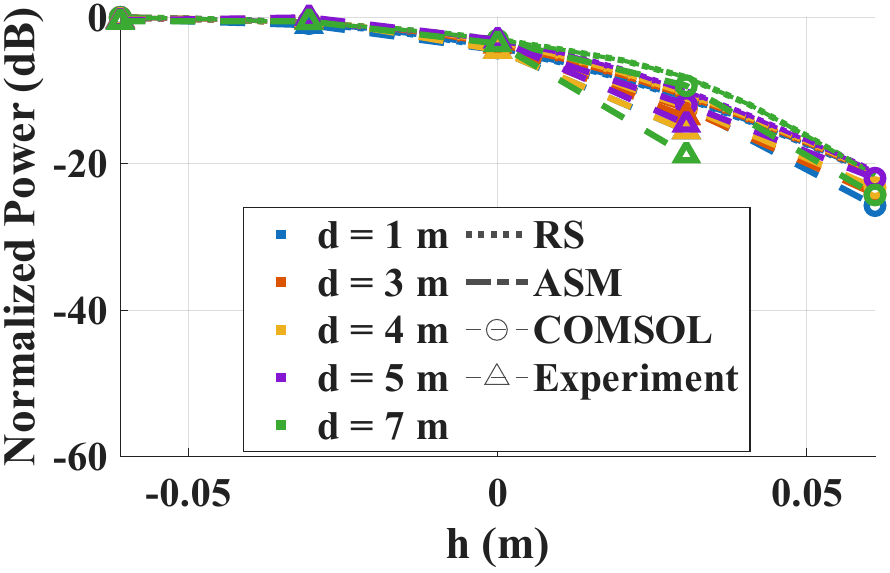}
        \caption{Normalized Power vs. $h$.}
        \label{fig:result_normalized_h}
    \end{subfigure}
    \hfill
    \begin{subfigure}{0.48\linewidth}
        \centering
        \includegraphics[width=\linewidth, trim=0 0 0 0, clip]{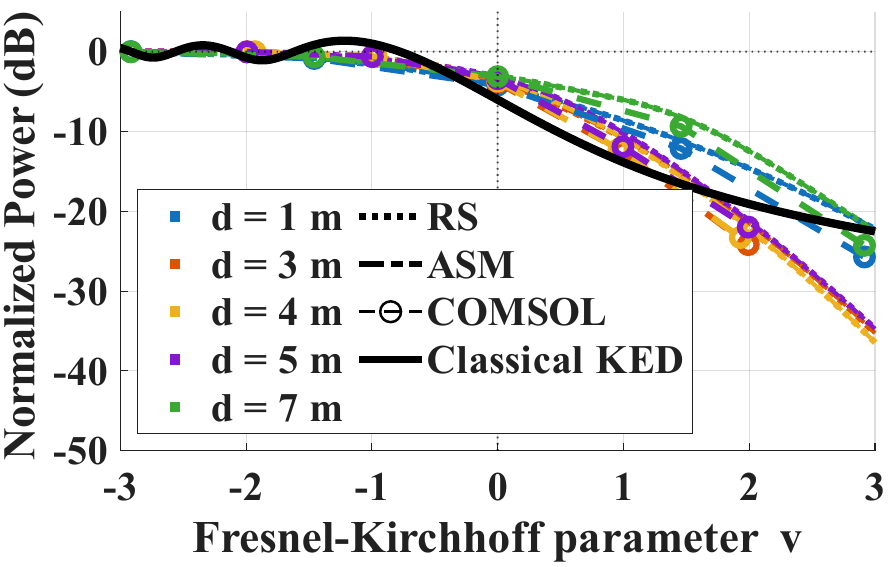}
        \caption{Normalized power vs. $v$.}
        \label{fig:Pr_v}
    \end{subfigure}
    \caption{Comparison of normalized received power versus (a) edge height $h$ from theoretical, simulation, and experimental results, and (b) Fresnel-Kirchhoff parameter $v$ between the THz finite-aperture near-field diffraction model and the classical KED model.}
    \vspace{-3mm}
    \label{fig:Pr_analysis}
\end{figure}

The experimentally captured signal strength after diffraction is illustrated in Fig.~\ref{fig:result_normalized_h}, where it shows the normalized received power compared with theoretical values and simulation results. It is evident that when the blockage is positioned perpendicularly to the axial distance, the received power exhibits a similar pattern for blockages located at different distances. As the vertical height \( h \) of the blockage increases, the received power diminishes. 
Furthermore, we observe that the normalized power results from experiments, theory, and simulations follow the same trend, which aids us in developing the localization mechanism described in Sec. ~\ref{sec:sensing}. In the experimental results, we did not provide the power received for full blockage, as it would not be reliably measurable due to the dominance of noise.

Additionally, we plot the normalized received power as a function of the Fresnel-Kirchhoff parameter \( v \), comparing our THz near-field finite-aperture diffraction model with the classical KED model~\cite{vavilov2019modeling} in Fig.~\ref{fig:Pr_v}. 
Unlike the universal curve of the classical model determined by the phase difference captured by $v$, the proposed model produces distinct curves for different blockage locations and shows notable differences in both the initial and tail regions. This difference arises because finite-aperture near-field diffraction depends on blockage location rather than a single phase parameter, as different blockage locations truncate different portions of the beam under a non-uniform field distribution. 
Moreover, the collimated beam exhibits diffraction characteristics different from the divergent illumination assumed in the classical model. Therefore, the classical KED model fails to capture diffraction behavior in the THz near-field finite-aperture regime.


\section{Parametric Inverse Diffraction Model for Blockage Localization}
\label{sec:sensing}
\begin{figure*}[t]
    \centering
    \begin{subfigure}[t]{0.23\linewidth}
        \centering
        \includegraphics[height=3.4cm]{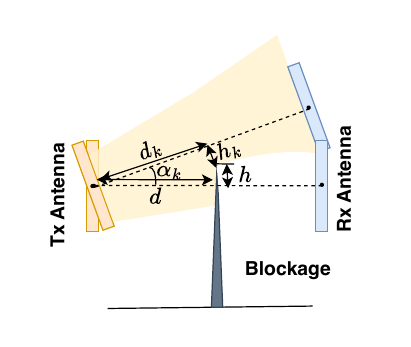}
        \caption{Rotation geometry.}
        \label{fig:rotation_geometry}
    \end{subfigure}\hfill
    \begin{subfigure}[t]{0.24\linewidth}
        \centering
        \includegraphics[height=3cm]{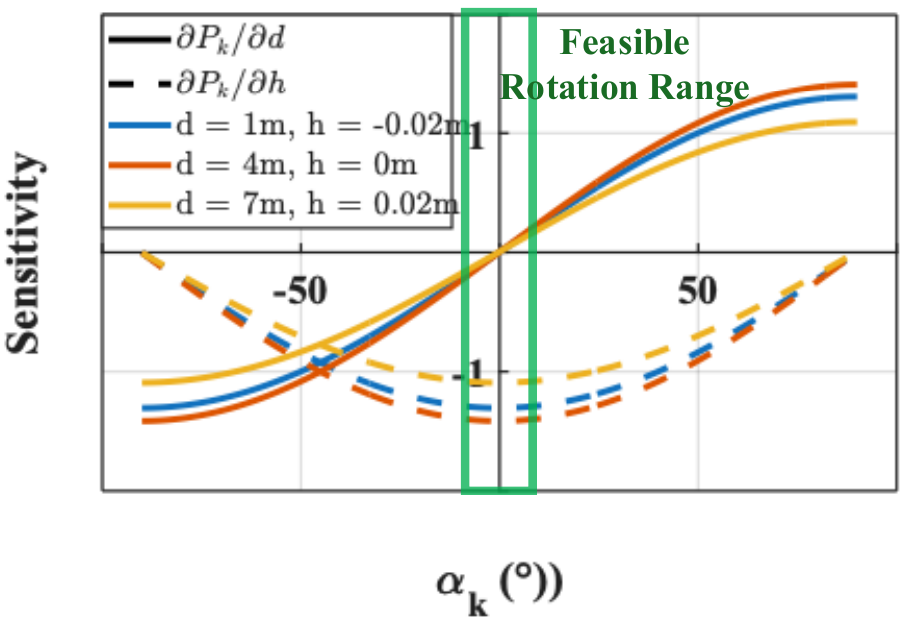}
        \caption{Sensitivity under rotation.}
        \label{fig:rotation_sensitivity}
    \end{subfigure}\hfill
    \begin{subfigure}[t]{0.24\linewidth}
        \centering
        \includegraphics[height=3cm]{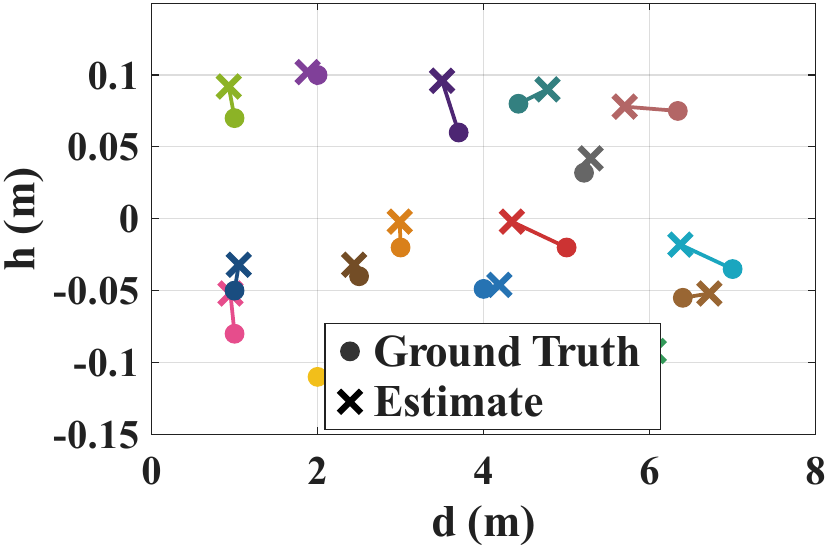}
        \caption{Localization results.}
        \label{fig:sensing}
    \end{subfigure}\hfill
    \begin{subfigure}[t]{0.24\linewidth}
        \centering
        \includegraphics[height=3cm]{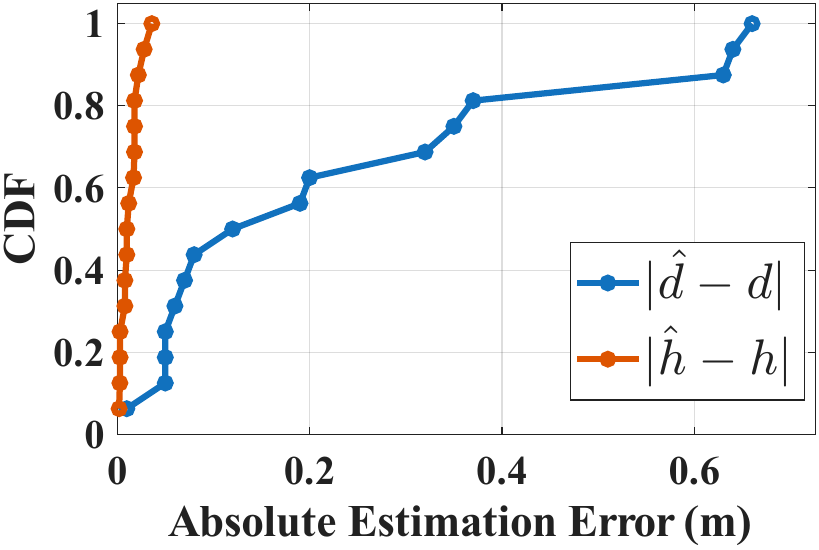}
        \caption{CDF of estimation error.}
        \label{fig:error_cdf}
    \end{subfigure}
    \caption{Blockage localization under the rotation configuration, where (a) depicts the rotation geometry, (b) plots the sensitivity of received power to axial distance $d$ and edge height $h$ versus rotation angle $\alpha_k$, (c) compares estimated blockage locations with ground truth, and (d) gives the CDF of absolute estimation errors for $d$ and $h$.
    }
    \vspace{-3mm}
    \label{fig:rotation_inverse_model}
\end{figure*}

While the theoretical diffraction model establishes a mapping from the blockage location $(d,h)$ to the received power $P_r$, localizing the blockage from power measurements is challenging due to the weak sensitivity of $P_r$ to $d$, as shown in Fig.~\ref{fig:sensitivity_analysis} and consistently observed in both COMSOL simulation and lab experiments.  
To overcome this limitation, we propose a rotation-based inverse diffraction model that enables localization of an unknown blockage.
\subsection{Rotation-based Parametric Model}


Consider a measurement setup where the Tx-Rx link is rotated 
by an angle $\alpha_k$ with respect to the original configuration while keeping the blockage fixed at location $(d,h)$, as shown in Fig. \ref{fig:rotation_geometry}. 
Under this rotation, the blockage location becomes
\begin{align}
    &d_k = d\cos{\alpha_k}+h\sin{\alpha_k}\label{eq:d_k},\\
    &h_k = -d\sin{\alpha_k}+h\cos{\alpha_k}.\label{eq:h_k}
\end{align}
The theoretical received power is therefore given by
\begin{equation}
    P_k(d,h)=f(d_k,h_k),
    \label{eq:P_k}
\end{equation}
where $f(\cdot)$ denotes the forward diffraction model in \eqref{eq:Pr}. For the same blockage, each rotation produces a different beam truncation and thus a distinct diffraction response, augmenting   
the information available for blockage localization.

Let $P_{\text{meas},k}$ denote the measured power under the $k$-th rotation configuration, which can be written as
\begin{equation}
    P_{\text{meas},k}=P_k(d,h) +\epsilon_k,
\end{equation}
where $\epsilon_k$ represents measurement noise. Given measurements from $N$ rotation configurations, the blockage location is estimated by solving the non-linear least-squares problem
\begin{equation}
(\hat d,\hat h)=
\arg\min_{d,h}
\sum_{k=1}^{N}
\left(
P_{\text{meas},k} - P_k(d,h)
\right)^2.
\end{equation} 
\subsection{Sensitivity Improvement via Rotation}
Rotation improves the observability of $d$ through its contribution to the more sensitive parameter $h_k$ in \eqref{eq:h_k}. Under the rotation configuration, applying the chain rule to \eqref{eq:P_k} with \eqref{eq:d_k} and \eqref{eq:h_k}, the sensitivities of the received power to the blockage location are given by:
\begin{align}
\frac{\partial P_k}{\partial d} &= f_d(d_k,h_k)\cos{\alpha_k}-f_h(d_k,h_k)\sin{\alpha_k},\\
    \frac{\partial P_k}{\partial h} &= f_d(d_k,h_k)\sin{\alpha_k}+f_h(d_k,h_k)\cos{\alpha_k}.
\end{align}
As shown in Sec. \ref{sec:forward_model_analysis}, the unrotated system satisfies  $|f_h|\gg|f_d|$, leading to
\begin{equation}
    \frac{\partial P_k}{\partial d} \approx -f_h(d_k,h_k)\sin{\alpha_k}.
\end{equation}
Thus, rotation injects part of the strong h-sensitivity into the sensitivity to $d$, providing additional information about $d$ and enabling joint estimation of the parameters $h$ and $d$.

The sensitivity variation with the rotation angle is shown in Fig.~\ref{fig:rotation_sensitivity}. The feasible rotation range is determined by the beam waist and corresponds to the region where the diffraction affects the received power. Outside the range, the received power becomes weakly dependent on the blockage location because diffraction becomes negligible when the blockage moves outside the beam illumination region. Since the transmitted Gaussian beam is highly collimated and propagates under the paraxial approximation, 
 this range is limited to several degrees for typical centimeter-scale apertures over meter-scale propagation distances. In this range, the sensitivity
$\frac{\partial P_k}{\partial d}$ increases approximately linearly with $\alpha_k$, and its magnitude grows with $|\alpha_k|$, as shown by the solid curve, indicating that larger rotations improve the observability of the axial distance $d$. Despite this improvement, the sensitivity $\frac{\partial P_k}{\partial h}$ remains larger than $\frac{\partial P_k}{\partial d}$, as indicated by the dashed curve.


\subsection{Inverse Model Analysis}
We emulate the localization process using $N=5$ rotation configurations and use the received power from COMSOL simulations as the measurement $P_{\text{meas},k}$. Due to the limited number of simulation cases, the selected rotation configurations may not correspond to the optimal set of angles.

The comparison between the estimated blockage locations and the ground truth is shown in Fig. \ref{fig:sensing}. Even with limited configurations, the estimates remain close to the true blockage locations, with errors generally within a few centimeters for $h$ and several tens of centimeters for $d$.  Fig. \ref{fig:error_cdf} further illustrates the estimation accuracy using the cumulative distribution function (CDF) of the absolute error. The absolute estimation error of $h$ is concentrated below $0.04$ m, while the maximum error of $d$ extends to $0.66$ m, indicating higher estimation accuracy for $h$, which is consistent with the sensitivity analysis under rotation configurations. These results demonstrate that the rotation-based parametric model captures sufficient geometric information for reliable blockage localization, providing the prior blockage information required for curved beam design to mitigate blockage.

\section{Conclusion}
We develop an analytical diffraction model for finite-aperture THz near-field propagation based on the Rayleigh-Sommerfeld integral and validate it through both simulations and experiments. The model reveals a strong asymmetry in the sensitivity of the received power to the edge height and axial distance, which limits direct power-based blockage localization and motivates a rotation-based sensing framework. 
 This framework improves the observability of the axial distance,  enabling reliable low-complexity blockage localization and providing useful insights for curved beam design to improve link robustness. As a first proof of concept, this work demonstrates the feasibility of power-based blockage localization using multiple diffraction responses obtained under rotation configurations. Future work will explore richer channel features for localization, practical measurement configurations beyond rotation, and more complex propagation environments.

\section*{Acknowledgment}
This work is supported by the U.S. National Science Foundation under grants CNS-2216332, CNS-2339811, CNS-2403286, and CNS-2451446, along with WINGS Center at SUNY Polytechnic Institute, AFRL (Award number 105882), as well as student Pin Chi Liang for data capture.

\bibliographystyle{ieeetr}
\bibliography{references}

\end{document}